\documentclass{article}
\usepackage{authblk}
\usepackage{amsmath}%
\usepackage{amsfonts}%
\usepackage{amssymb}%
\usepackage{graphicx}
\usepackage[margin=1in]{geometry}

\title{Grazing incidence x-ray diffraction studies of lipid-peptide mixed monolayers during shear flow}

\author[1]{Pradip Kumar Bera}
\author[1,2]{Ajoy Kumar Kandar}
\author[1,3]{Rema Krishnaswamy}
\author[4]{Philippe Fontaine}
\author[5]{Marianne Imp\'eror-Clerc}
\author[5]{Brigitte Pansu}
\author[5]{Doru Constantin}
\author[6]{Santanu Maiti}
\author[6]{Milan K. Sanyal}
\author[1,*]{A.K. Sood}

\affil[1]{Department of Physics, Indian Institute of Science, Bangalore 560012, India}
\affil[2]{Present address: Soft Condensed Matter, Debye Institute for Nanomaterials Science, Utrecht
University, Princetonplein 1, 3584 CC Utrecht, The Netherlands}
\affil[3]{Present address: School of Liberal Studies, Azim Premji University, Bangalore, 560100, India}
\affil[4]{SOLEIL Synchrotron, L'Orme des Merisiers, Saint-Aubin - BP48, 91192 GIF-sur-YVETTE CEDEX, France}
\affil[5]{Laboratoire de Physique des Solides, Unit\'e Mixte de Recherche 8502 Centre National de la Recherche Scientifique, Universit\'e Paris-Sud 11, 91405 Orsay Cedex, France}
\affil[6]{Saha Institute of Nuclear Physics, 1/AF, Bidhannagar, Kolkata-700064, India}

\affil[*]{Email: asood@iisc.ac.in}

\begin{document}
\date{}
\maketitle

\begin{abstract}
Grazing Incidence X-ray Diffraction (GIXD) studies of monolayers of biomolecules at the air-water interface give quantitative information of in-plane packing, coherence lengths of the ordered diffracting crystalline domains and the orientation of hydrocarbon chains. Rheo-GIXD measurements revel quantitative changes in the monolayer under shear. Here we report GIXD studies of monolayers of Alamethicin peptide, DPPC lipid and their mixtures at the air-water interface under the application of steady shear stresses. The Alamethicin monolayer and the mixed monolayer show flow jamming transition. On the other hand, pure DPPC monolayer under the constant stress flows steadily with a notable enhancement of area/molecule, coherence length, and the tilt angle with increasing stress, suggesting fusion of nano-crystallites during flow. The DPPC-Alamethicin mixed monolayer shows no significant change in the area/DPPC molecule or in the DPPC chain tilt but the coherence length of both phases (DPPC and Alamethicin) increases suggesting that the crystallites of individual phases are merging to bigger size promoting more separation of phases in the system during flow. Our results show that Rheo-GIXD has the potential to explore in-situ molecular structural changes under rheological conditions for a diverse range of confined biomolecules at the interfaces.

\end{abstract}

\section*{Introduction}
Langmuir monolayer, a molecularly thin film of amphiphilic molecules stabilized on a liquid-air interface, is an important model system for studying self-organized biological structures such as cell membranes, lung alveoli, and also has important industrial applications like in foam, emulsion, etc \cite{als1994principles,kaganer1999structure,fuller2012complex,bera2019experimental}. A combination of Grazing Incidence X-ray Diffraction (GIXD), specular x-ray reflectivity (XR) and more recently electrochemical scanning tunneling microscopy (EC-STM) of Langmuir-Blodgett (LB) monolayers have been used to understand different kinds of phase transitions, molecular structure and orientation within crystalline domains (crystallites), formation of single layer and bi-layers \cite{kjaer1987ordering,kjaer1994some,miller2008probing,maiti2018structural,maiti2018understanding}. Mixed systems like lipid-cholesterol and lipid-peptide monolayers have been studied to probe the interactions of lipids with other molecules and their relative orientation \cite{watkins2009structure,wu2005interaction,neville2008protegrin,ivankin2010cholesterol,watkins2011membrane,broniatowski2016studies}.

Alamethicin is an antimicrobial peptide, produced by many living organisms to defend against gram-negative and gram-positive bacteria, fungi, enveloped viruses, eukaryotic parasites, and even tumor cells. Alamethicin isolated from $Trichoderma\:viride$ has 20 residue peptides with predominantly $\alpha$-helical structure. In the helical conformation, the length of the molecule is 33 \AA. The helix oriented parallel to the interface is called the surface (S) state. If it is inserted into the lipid matrix with the helical axis perpendicular to the interface, it is called the inserted (I) state. The aggregation properties and flow behavior of Alamethicin in the form of Langmuir monolayer were studied using fluorescence microscopy and surface rheology \cite{krishnaswamy2008aggregation}. Fluorescence microscopy showed the coexistence of liquid-expanded and solid phases. The net area fraction of the solid phase increases with concentration. Interfacial rheology showed that the peptide monolayer at concentration 800 \AA$^2$/molecule and above has yield stress which increases with surface concentration.

Biological lipid rafts are dynamic self-organized membrane microdomains that can recruit specific peptides and lipids selectively while excluding others \cite{edidin2003state}. The lipid DPPC shows a variety of different ordered states due to the steric and van der Waals interactions between neighboring head groups and alkyl chains. DPPC monolayers exhibit a disordered liquid-expanded (LE) phase that transforms into a liquid-condensed (LC) phase with long-range orientational and short-range positional order at high concentration. The DPPC monolayer has been studied using in-situ fluorescence microscopy to correlate domain dynamics with shear flow \cite{espinosa2011shear,hermans2014interfacial,kim2011interfacial,choi2011active}. In the high concentration limit, the thin domain boundaries were only visible by fluorescence and it was proposed that the interlocked domains give rise to the yield stress response of the LC-DPPC monolayer. The domain topology was preserved for small shear rates. The lipid interaction with peptides and their structural organization are governed by electrostatic and hydrophobic interactions. Recently molecular imaging techniques like STM, surface-enhanced infrared absorption (SEIRA) spectroscopy, etc have revealed hexameric pore formation in the lipid membranes \cite{pieta2012direct,forbrig2018monitoring}. So far there is no structural study of the model membranes at air-water interface under shear, though in-situ GIXD has been proposed as a potential probe to monitor the dynamic properties of the crystallites of the model membranes \cite{kim2011interfacial,fuller2012complex}.

In this work, we present in-situ interfacial rheology along with GIXD to understand changes in the membrane lattice structure under the non-equilibrium steady-state flow condition. Rheo-GIXD measurements are done on the three model systems: Alamethicin, DPPC, and DPPC-Alamethicin mixed monolayers, at different applied stress.

\section*{Experimental Details}

\subsection*{Materials}
The lipid with two hydrocarbon chains $1,2 \mbox{-} dipalmitoyl \mbox{-} sn \mbox{-} glycero \mbox{-} 3 \mbox{-} phosphocholine$ (DPPC) and the peptide Alamethicin (all from M/s Avanti Polar Lipids, Inc.) were used without further purification. A mixture of chloroform and methanol (1:1 v/v) was used as a volatile solvent to dissolve the peptide and lipid molecules. The required amount of the solution was spread on the air-water interface using a microsyringe (M/s Hamilton, 50 $\mu$L) to obtain the interfacial layer between the bi-cone and the co-centric homemade shear cell, after the evaporation of the solvent \cite{krishnaswamy2008aggregation}. A deionized water sub-phase (M/s Millipore, with a resistivity of 18.2 M$\Omega$.cm) was used for the DPPC monolayer. For pure Alamethicin and DPPC-Alamethicin (molar ratio [Alamethicin]/[DPPC] = 1:2) mixed monolayers, the sub-phase was the aqueous solution of 0.1 mole NaCl (pH 7), which was adjusted with 10$^{-3}$ mole phosphate buffer (Na$_{2}$HPO$_{4}$:NaH$_{2}$PO$_{4}$ 1:1, M/s Merck).

\subsection*{Rheo-GIXD measurement}

\begin{figure}
\centering
\includegraphics[width=1.0\textwidth]{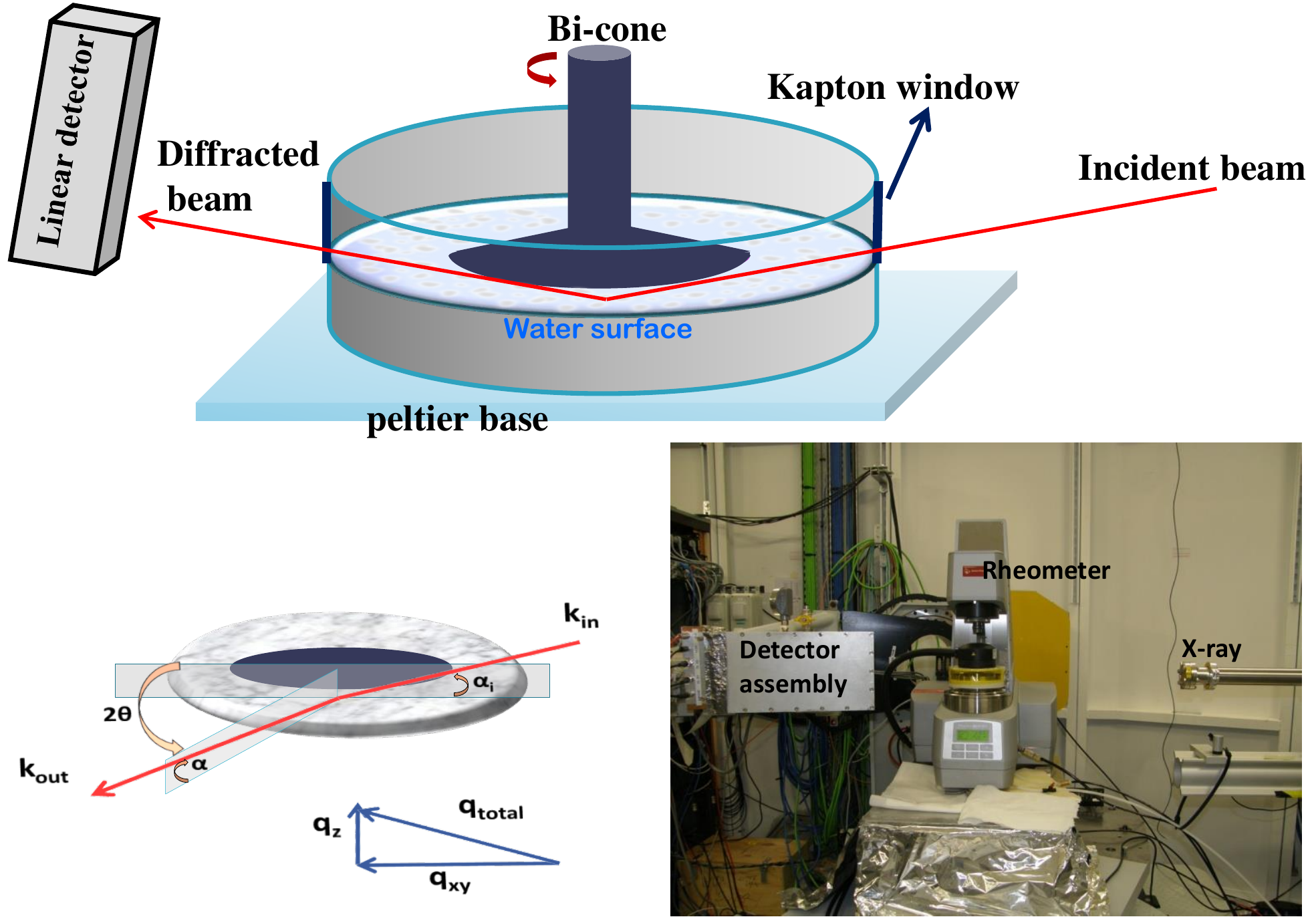}
\caption{(Color online) schematic of the in-situ Rheo-GIXD setup, showing the water-filled IRS cell on the rheometer's peltier base, the position of the bi-cone on the interface and the path of the x-ray beam through the Kapton window striking the annular shaped interface (top). (Bottom left) schematic of the GIXD mechanism: the vertical incidence angle ($\alpha_i$), the horizontal scattering angle ($2\theta$) and the vertical exit angle ($\alpha$); in-plane wave vector q$_{xy} \simeq (4\pi/\lambda) \sin{(2\theta/2)}$ and out-of-plane wave vector q$_{z}=(2\pi/\lambda)(\sin{\alpha} + \sin{\alpha_i})$ are shown. (Bottom right) photograph of the experimental setup showing the x-ray source, the rheometer on a z-stage and the detector assembly attached to the goniometer.}
\label{F1}
\end{figure}

We have calibrated our rheo-GIXD set up using behenic acid monolayer as a test sample. The 2D diffraction pattern (see Supporting Information Section A) is matching with literature \cite{pignat2006ph} and validates our GIXD setup. Also, the sensitivity of the monolayers to the small imposed torque on the measuring bi-cone geometry was checked using test monolayers with cholesterol, which is known for showing very low surface viscosity \cite{evans1980surface} (see Supporting Information Section B). With the cholesterol and cholesterol mixed monolayers we get very high value of the shear rate ($\dot{\gamma}$) which confirms the good sensitivity of the monolayers even to the very small interfacial stress ($\sigma$ in units of $\mu$Pa-m) imposed by the rheometer.

After spreading the solution at 300 K the cell was covered by Teflon cover and then waited for 2000 s to let the spreading solvent evaporate under the slow helium flow. During this, an oscillatory shear of strain amplitude $\gamma_0$ = 0.001 with an angular frequency $\omega$ = 10 rad/s was applied to follow up the formation of the monolayer. To maintain identical initial conditions before each creep measurement, monolayers were presheared at $\sigma$ = 250 $\mu$Pa-m for 200 s and then allowed the system to equilibrate for 300 s. After 500 s from the starting of creep measurements, GIXD measurements were started to scan the system in the steady flow state. The rheo-GIXD experiments were carried out at the SIRIUS beamline of the SOLEIL Synchrotron, France using an x-ray photon energy of 8 keV ($\lambda \approx$ 1.55 \AA) at 285 K \cite{fontaine2014soft}. This low value of temperature is chosen to minimize the evaporation of water which could change the bicone coupling to the interface. A stress-controlled rheometer (M/s Anton Paar, model MCR-501) fitted with a homemade interfacial shear cell (radius = 65 mm) based on the bi-cone geometry (radius = 34.14 mm) was mounted on the SIRIUS beamline. A schematic of the experimental setup is shown in Figure \ref{F1}. The dimension of the x-ray beam footprint on the liquid surface was maintained to be $\sim$ 1.5 mm $\times$ 20 mm (velocity-gradient velocity direction) by the slits attached to the x-ray source. As the x-ray grazing angle is very small, the shear cell was slightly overfilled to get a just inverted meniscus. The position of the rheometer was set to have the x-ray beam $\sim$ 5 mm away from the cone edge. After each load, to make the x-ray footprint strike at the same position on the surface, the height of the motorized stage was adjusted in order to bring the liquid surface to the desired height by scanning the specular reflection of the x-ray. Note, the local velocity of the region being scanned is $\approx \dot{\gamma} \times$ $y$ (note $\dot{\gamma}$ is the global shear rate in the system), $y=$ 25 mm; is the distance of the x-ray footprint from the cell wall. Water-saturated helium was injected slowly inside the cell from the top to reduce scattering from the air. The monochromatic x-ray beam was adjusted to strike the interface at an incident angle $\alpha_i$ = 2.28 mrad, which corresponds to 0.85 $\alpha_c$, where $\alpha_c$ is the critical angle of air-water interface \cite{als1994principles} corresponding to the wavelength. The linear (1D) gas-filled position sensitive detector (PSD) fitted with the goniometer was used to record the diffraction pattern by varying the horizontal angle $2\theta$ from low to high. Soller slits with angular resolution 0.02 degree were used.

\subsection*{GIXD Data analysis}
Two dimensional (2D) diffraction plots for all the three monolayers at rest are shown in Figure \ref{F2}. As a check, a smooth background is observed in GIXD from the clean water surface without any feature. The in-plane scattering wave vector q$_{xy}$ gives information about Bragg peaks in the velocity-velocity gradient plane (V$\times \nabla$V). On the other hand, the out-of-plane scattering wave vector q$_{z}$ gives information about the Bragg rods \cite{watkins2011membrane,als1994principles}. q$_{xy}$ and q$_{z}$ are expressed in terms of vertical incidence angle ($\alpha_i$), horizontal scattering angle ($2\theta$) and vertical exit angle ($\alpha$) as \cite{kjaer1994some}:

\begin{equation} \label{eqnqxy}
\begin{split}
\frac{q_{xy}}{k} & = \sqrt{(\cos{\alpha} - \cos{\alpha_i})^2 + 2\cos{\alpha}\cos{\alpha_i}(1 - \cos{2\theta})} \\
								& \simeq \sqrt{(1 + \cos{\alpha}^2 - 2\cos{\alpha}\cos{2\theta})} \\
								& \simeq 2 \sin{(2\theta/2)} + O(\alpha^2) \\
\end{split}
\end{equation}
\begin{equation} \label{eqnqz}
\frac{q_{z}}{k} = \sin{\alpha} + \sin{\alpha_i} \\
\end{equation}

Where $k = 2 \pi/\lambda$ and $\cos{\alpha_i} \approx 1$ for very small value of $\alpha_i$.

The observed peaks are well separated in the q$_{xy}$-q$_{z}$ contour plots. We note that the relatively more noise in the data compared to the monolayers prepared in Langmuir trough is expected because our experiments are done on a spread monolayer in place of compressing it from a liquid expanded phase and later, it is in the flow state. \cite{neville2008protegrin,watkins2009structure,watkins2011membrane}. We have adopted the box integration method for each peak as discussed below. Bragg peaks are observed by integrating the contours from q$_{z}$ = 0 \AA$^{-1} $ to 0.1 \AA$^{-1}$ and from 0.3 \AA$^{-1}$ to 0.5 \AA$^{-1}$. Bragg peaks are fitted with Voigt function along with background intensity to get the peak centers and the full width at half maximum (FWHM) \cite{watkins2009structure}. For DPPC, lattice distance d$_{hk}$ = 2$\pi$/q$_{hk}$ are extracted using Bragg peaks q$_{02}$ and q$_{11}$ and then fitted to 2D centered rectangular unit cell model to get the lattice parameters a and b \cite{watkins2011membrane,kjaer1994some} and hence area/molecule. With the Scherrer equation, FWHM of the Bragg peaks were used to determine the coherence length L (L= 2$\pi$/FWHM) which can be approximated as the average size of the nano-crystallites.

Bragg rod profiles are obtained by integrating the contours for $\Delta$q$_{xy}$ = q$_{02}\pm 0.015$ \AA$^{-1}$ (near q$_{02}$), and q$_{11}\pm 0.030$ \AA$^{-1}$ (near q$_{11}$) (see Figure \ref{F2}). In our GIXD data q$_{02}$ and q$_{11}$ peaks have long tails in upward and downward directions respectively, suggesting large variation in the tilt angle of hydrocarbon chains with respect to the interface normal towards the nearest-neighbor direction (NN). The maximum tilt angle ($\delta$) is calculated using $\delta$ = $\Delta$q$_z$/(2$\pi$/a), where $\Delta$q$_z$ is the peak to peak distance in the I vs q$_z$ plot \cite{watkins2009structure}. We have restricted our study for q$_{xy} \geq 1.0$ \AA$^{-1}$, below this limit, the noise increases significantly towards the direct beam.

\section*{Results and discussion}

\subsection*{Equilibrium study of Alamethicin, DPPC, DPPC-Alamethicin mixed monolayer}

\begin{figure}
\centering
\includegraphics[width=1.0\textwidth]{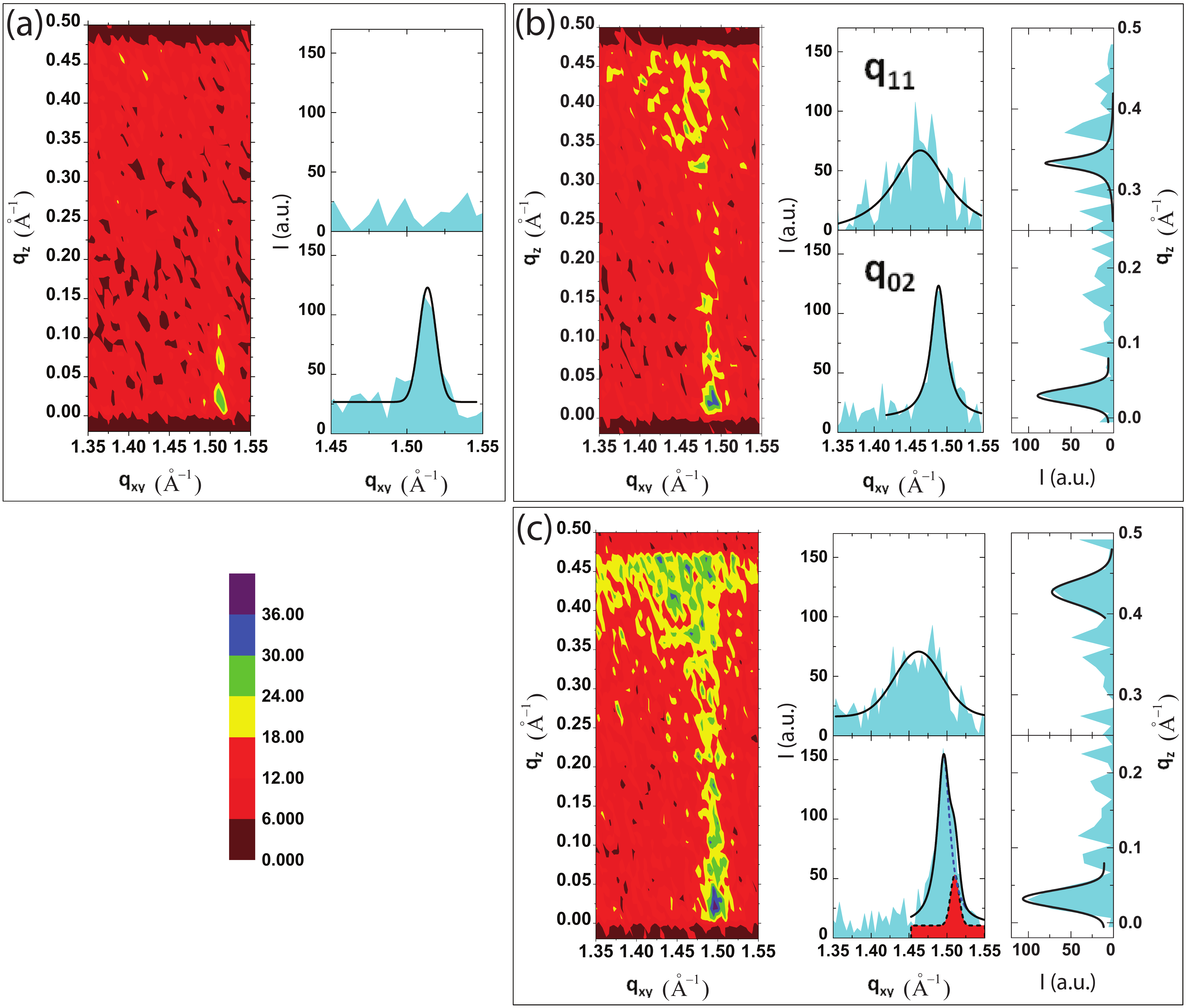}
\caption{(Color online) GIXD intensity contours in (q$_{xy}$, q$_{z}$) plane, Bragg peaks (I vs q$_{xy}$) and Bragg rod profiles (I vs q$_{z}$) of the three monolayers (a) Alamethicin (b) DPPC and (c) DPPC-Alamethicin are shown under no shear condition at 285 K. Solid lines are fits using Voigt function. In (c) for the bottom Bragg peak, the solid line is the resultant fit with two peaks (blue dotted line and red shaded black dotted line). Color bars represent intensity values in contours.}
\label{F2}
\end{figure}

Before applying shear to the monolayers at the annular-shaped air-water interface between the bi-cone and the shear cell, their structural properties were characterized. Figure \ref{F2} shows the equilibrium diffraction patterns of Alamethicin, DPPC, DPPC-Alamethicin mixed monolayer. The Alamethicin monolayer was prepared for 12 \AA$^2$/molecule surface concentration as lower concentrations do not give rise to a measurable diffraction peak in the GIXD. The equilibrium GIXD pattern shows up a strong peak at q$_{xy}$ = 1.514 \AA$^{-1}$ near q$_{z}$ = 0 confirming that the Alamethicin molecules are adsorbed on the surface. The observed strong peak due to the Alamethicin corresponds to the pitch of the helix of 4.15 \AA (Figure \ref{F4}c) which is quite small compared with the pitch of 5.4 \AA for a free $\alpha$-helix. This reduction in helix pitch is due to the compact packing of Alamethicin molecules on the water surface at this high concentration, consistent with the previous study of the helical scattering distribution of Alamethicin \cite{spaar2004conformation}. The coherence length estimated from the measured linewidth ($\sim$ 475 \AA) suggests that there are domains of at least 14 correlated molecules. The expected hexagonal lattice ordering, forming holes inside these domains \cite{pieta2012direct}, with lattice parameters of a = 19 \AA, should show a Bragg peak in the low q range which is not seen in our experiments due to high background intensity near the direct beam, and hence we cannot estimate the area/molecule from the GIXD data.

The GIXD pattern from DPPC (solution concentration of 0.5 mg/mL) shown in Figure \ref{F2}b gives area/molecule = 42.1 \AA$^2$/molecule. DPPC has the 2D ordering of molecules on the water surface and gives rise to two well-separated two Bragg peaks (Figure \ref{F2}b) at q$_{xy}$ = 1.464 \AA$^{-1}$ (q$_{z}$ = 0.43 \AA$^{-1}$) and q$_{xy}$ = 1.489 \AA$^{-1}$ (q$_{z}$ = 0.03 \AA$^{-1}$). The relative intensity of these two peaks is $\sim$ 2:1 as expected for the DPPC monolayer \cite{watkins2009structure}. The diffraction pattern is analyzed with the centered-rectangular unit cell model of rod-shaped alkyl chains with uniform NN tilts \cite{watkins2009structure} (Table \ref{T1}). The coherence length and the hydrocarbon chain tilt angle are consistent with the previous studies \cite{watkins2009structure}.

The DPPC-Alamethicin mixed monolayer was prepared with a molar ratio of 1:2 and with surface concentrations 12 \AA$^2$/Alamethicin-molecule. The GIXD clearly shows three Bragg peaks (Figure \ref{F2}c), one is at 1.510 \AA$^{-1}$ representing Alamethicin helix pitch and the other two at 1.463 \AA$^{-1}$ and 1.496 \AA$^{-1}$ with 2:1 intensity ratio, associated with the DPPC molecular ordering in the monolayer. The estimated area/molecule of DPPC is 42.0 \AA$^2$/molecule which is very close to the pure DPPC monolayer (Table \ref{T2}). The hexagonal structure of Alamethicin in DPPC-Alamethicin \cite{pieta2012direct} mixture could not be observed due to the high direct-beam leakage intensity at low q$_{xy}$. Note that in equilibrium, the Alamethicin helix peak is on the shoulder of the DPPC q$_{02}$ Bragg peak, but with shear flow coherence lengths corresponding to the DPPC q$_{02}$ peak and the Alamethicin helix peak increase drastically and thus Alamethicin helix peak stands well separated in the GIXD pattern (see Figure \ref{F6}b).

\subsection*{Stress controlled flow curve}

\begin{figure}
\centering
\includegraphics[width=0.8\textwidth]{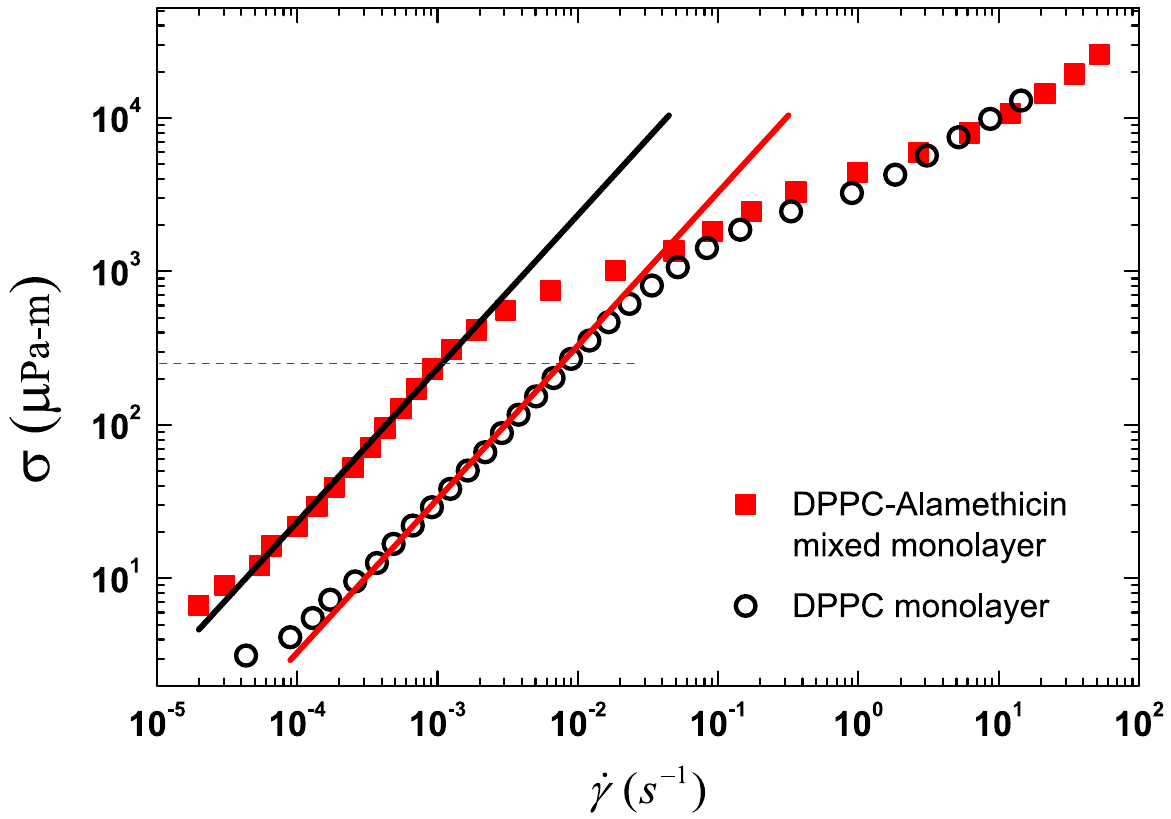}
\caption{Flow curve, shear stress ($\sigma$) versus the shear rate ($\dot{\gamma}$) obtained in the controlled shear stress (CSS) mode with a waiting time of $30 \:s$ for each data point, is shown for DPPC monolayer and DPPC-Alamethicin mixed monolayer at the air-water interface. Solid lines are of slope $\sim 1$. Dotted line is the approximate cutoff of the linear flow region ($\sim 250 \:\mu Pa\mbox{-}m$).}
\label{F3}
\end{figure}

For the flow curve and other rheological characterization of Alamethicin monolayers, see Ref. \cite{krishnaswamy2008aggregation}. Figure \ref{F3} shows the stress-controlled flow curves of DPPC, DPPC-Alamethicin mixed monolayers. The flow curves of the monolayers are very similar to the monolayer studied by Majumdar et al (see Figure 3 of Ref. [\cite{majumdar2011shear}]), where surface deformation profile is studied and the flow inhomogeneity or shear banding is reported in the non-linear region. In order to avoid flow inhomogeneity or shear banding, we have chosen the linear flow region as our working region for the two monolayers (as indicated by the blue and black lines with slope $\sim 1$). The approximate upper cutoff stress is the chosen pre-shear stress ($\sigma = 250 \:\mu Pa\mbox{-}m$) for each creep measurement to erase the history of the system (for details please see section Experimental Details).

\subsection*{Creep study of the Alamethicin monolayer}

\begin{figure}
\centering
\includegraphics[width=1.0\textwidth]{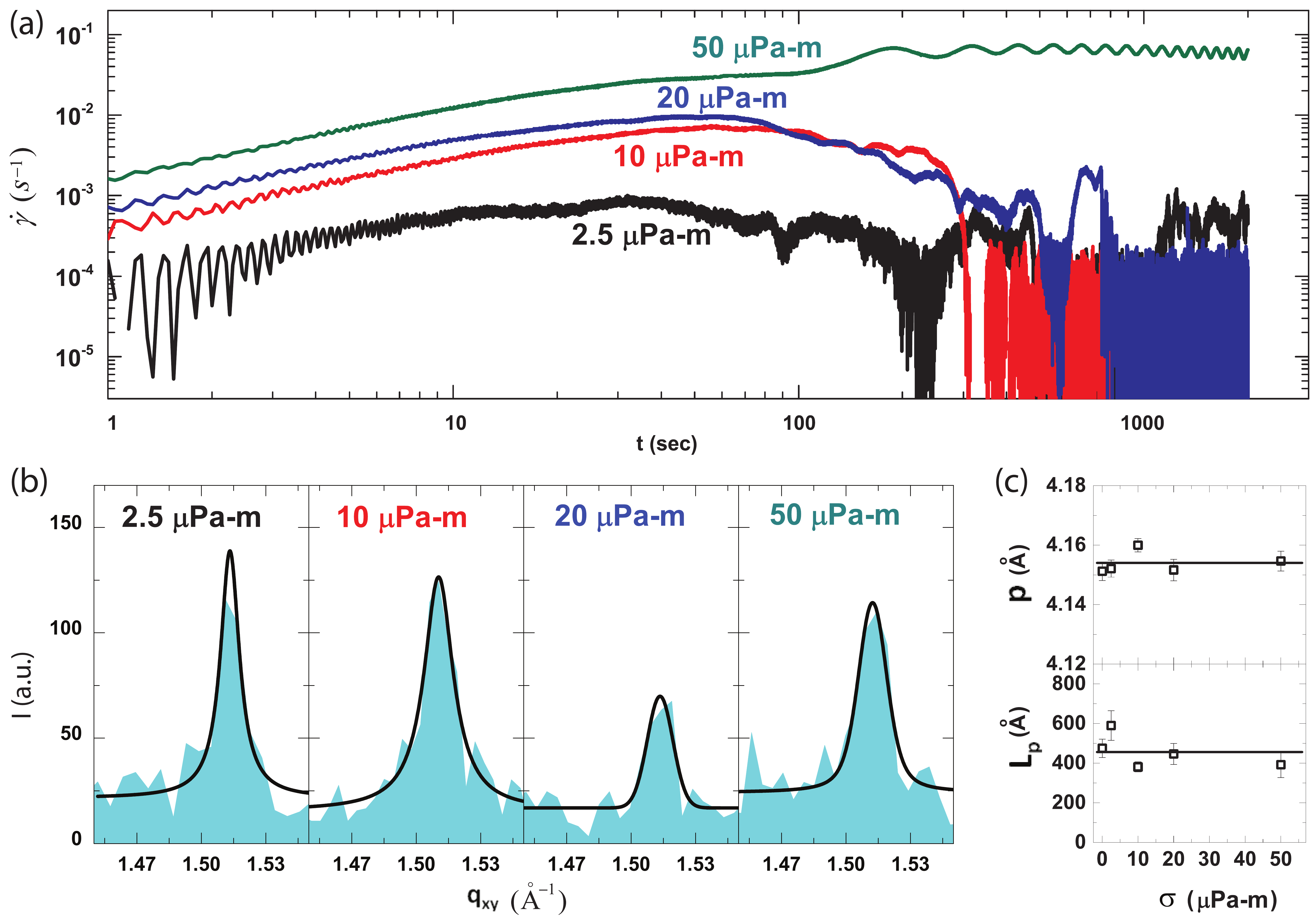}
\caption{(Color online) Rheo-GIXD creep data of the Alamethicin monolayer (presheared for 200 s followed by 300 s waiting before each measurement; see text): (a) creep curves; shear rate ($\dot{\gamma}$) vs time (t) (applied stress $\sigma$ is mentioned close to the curves), (b) Bragg peak (I vs q$_{xy}$) for different $\sigma$ are shown. Solid lines are fits using Voigt function. The Bragg peak corresponds to the helix pitch of Alamethicin. (c) Helix pitch (p) and coherence length (L$_p$) are plotted vs $\sigma$. Straight horizontal lines represent average values of p and L$_p$ respectively.}
\label{F4}
\end{figure}

We now proceed to examine the structural changes inside the monolayers in non-equilibrium steady-state, under different shear stress conditions. Figure \ref{F4}a shows the creep behavior of Alamethicin monolayer studied as a function of applied stress up to 50 $\mu$Pa-m. For all applied stresses ($\sigma$), shear rate ($\dot{\gamma}$) increases linearly with time for $\sim$ 60 s showing significant shear rejuvenation in the monolayer before going to the final steady-state. The stress values ranging from 2.5 $\mu$Pa-m to 50 $\mu$Pa-m are much above the stress resolution (0.3 $\mu$Pa-m) of the rheometer. For 2.5 $\mu$Pa-m $\leq \sigma \leq$ 20 $\mu$Pa-m, the shear rate is $\sim$ 10$^{-2}$ to 10$^{-4}$ s$^{-1}$ (much higher than the resolution $\sim$ 10$^{-7}$ s$^{-1}$). After $\sim$ 200 s, the shear rate decreases and fluctuates about zero, though with a positive value of the average shear rate. This observation of shear rate fluctuating about zero is seen in the stress-induced jamming behavior in bulk rheology of laponite clay suspension \cite{majumdar2012statistical}. The fluctuations seen in $\dot{\gamma}$ are genuine and their statistical properties (not discussed in the paper as it is outside the scope of this work) are similar to that in Ref. [ \cite{majumdar2012statistical}]. In comparison, at 50 $\mu$Pa-m, $\dot{\gamma}$ attains a steady-state value of $\sim$ 0.06 $s^{-1}$. Figure \ref{F4}b shows the GIXD data at four values of stress, integrated over time from 500 s to 2000 s. The helix peak position remains constant with increasing $\sigma$ but the line width shows variation reflecting the changes in the domain size (Figure \ref{F4}c). However, there is no systematic variation of the coherence length with applied stress.

\subsection*{Creep study of the DPPC monolayer}

\begin{figure}
\centering
\includegraphics[width=1.0\textwidth]{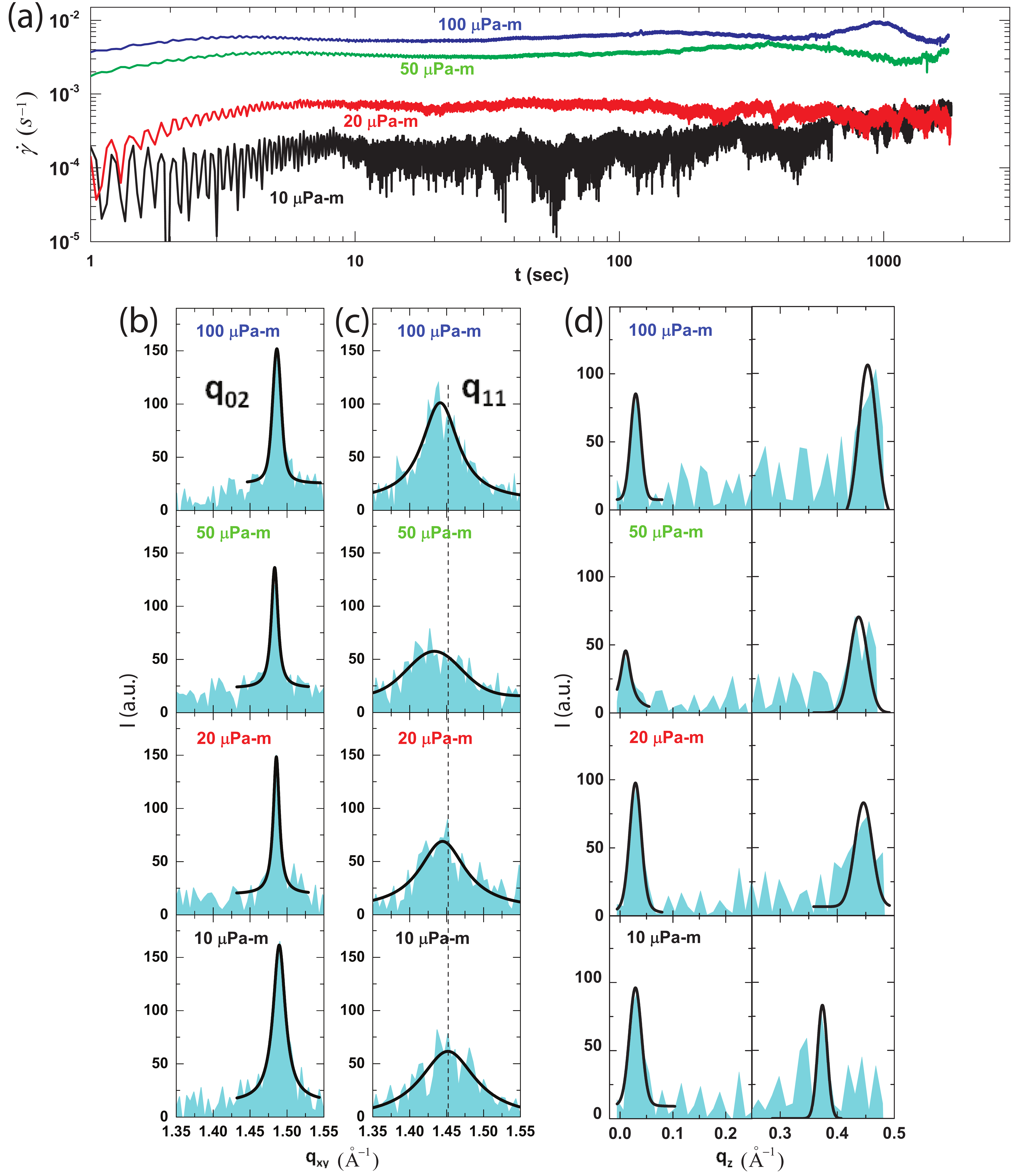}
\caption{(Color online) Rheo-GIXD creep data of the DPPC monolayer (presheared for 200 s followed by 300 s waiting before each measurement; see text): (a) creep curves; $\dot{\gamma}$ vs t, (b) Bragg peak q$_{02}$, (c) Bragg peak q$_{11}$, (d) Bragg rod profile for different $\sigma$ are shown. Solid lines in (b), (c), (d) are fits using Voigt function. Dashed vertical line in (c) has position q$_{xy}$ = 1.451 \AA$^{-1}$.}
\label{F5}
\end{figure}

The creep behavior of the DPPC monolayer was studied up to 100 $\mu$Pa-m (Figure \ref{F5}). Unlike Alamethicin monolayer, DPPC shows neither substantial shear rejuvenation nor flow jamming. For a given $\sigma$, the steady-state shear rate is an order of magnitude low compared to Alamethicin monolayer (50 $\mu$Pa-m data can be compared). The Bragg peaks q$_{02}$, q$_{11}$, and Bragg rod profiles are shown in Figure \ref{F5}b,c,d. The peak position of q$_{02}$ does not change with stress whereas q$_{11}$ peak position shifts to lower values, suggesting elongation of the unit cell under shear flow. Additionally, the width of q$_{02}$ decreases with increasing $\sigma$, suggesting the fusion of crystallites during flow. The tilt angle of the hydrocarbon chains increases with $\sigma$ (Figure \ref{F5}d). Table-\ref{T1} summarizes these results and are plotted in Figure \ref{F8}. The Rheo-GIXD data bring out that the DPPC crystallites' size increases under applied stress.

\subsection*{Creep study of the mixed monolayer}

\begin{figure}
\centering
\includegraphics[width=1.0\textwidth]{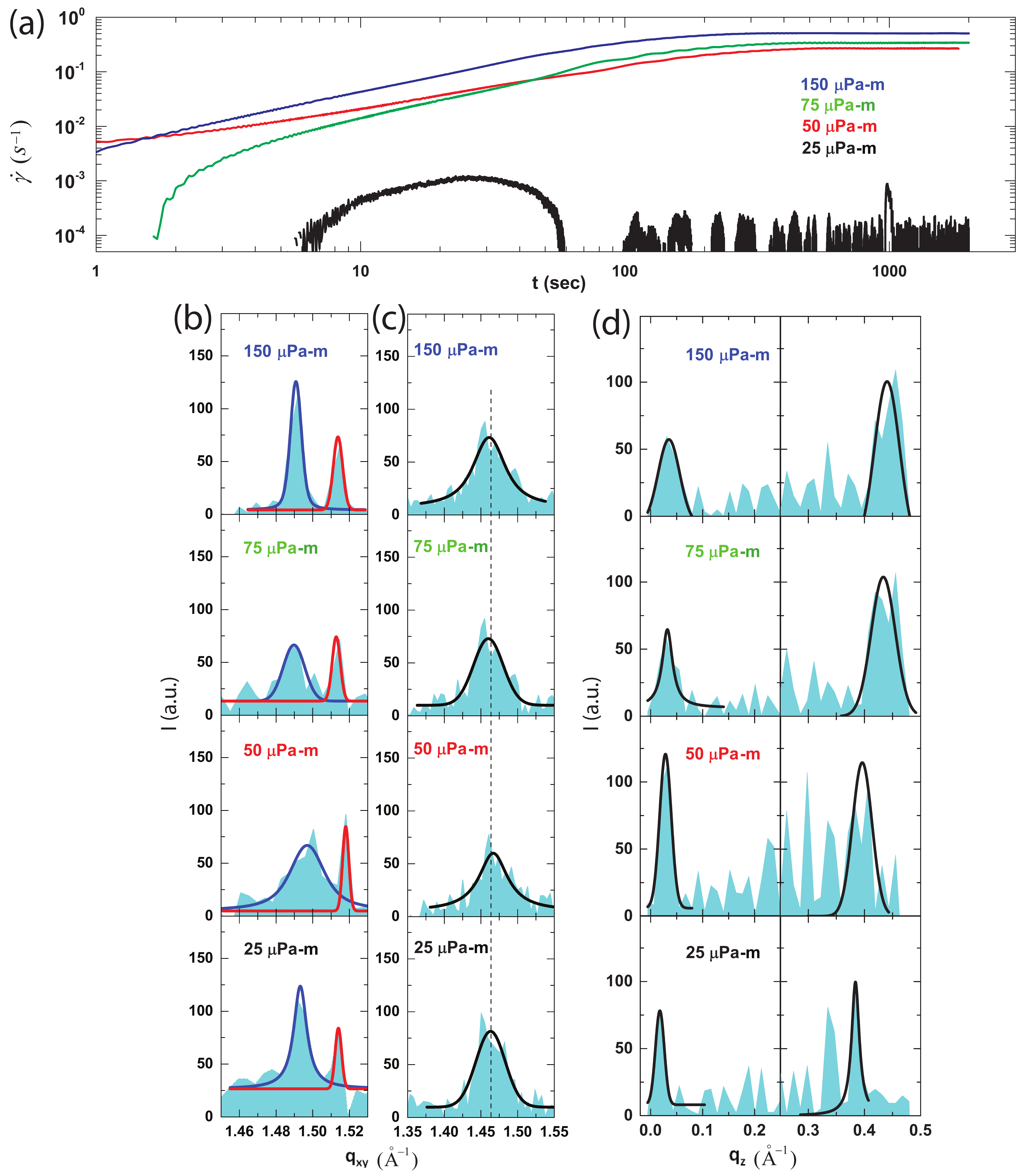}
\caption{(Color online) Rheo-GIXD creep data of the DPPC-Alamethicin mixed monolayer with molar ratio P/L=1/2 (presheared for 200 s followed by 300 s waiting before each measurement; see text): (a) creep curves; $\dot{\gamma}$ vs t, (b) Bragg peak q$_{02}$ (blue solid fit), Alamethicin helix peak (red solid fit), (c) Bragg peak q$_{11}$, (d) Bragg rod profile for different $\sigma$ are shown. Solid lines in (b), (c), (d) are fits using Voigt function. Dashed vertical line in (c) has position q$_{xy}$ = 1.463 \AA$^{-1}$.}
\label{F6}
\end{figure}

Figure \ref{F6} shows the creep behavior of the DPPC-Alamethicin mixed monolayer studied up to 150 $\mu$Pa-m. Shear rejuvenation is observed with $\dot{\gamma}$ increasing linearly with time. At 25 $\mu$Pa-m it shows rejuvenation up to 30 s and then goes to the flow jammed state after 60 s of flow similar to the pure Alamethicin monolayer. At 50 $\mu$Pa-m and above it goes to a steady flow state with an enhanced $\dot{\gamma}$ compared to pure Alamethicin monolayer, which is orders of magnitude higher compared to pure DPPC monolayer. This suggests that the DPPC crystalline domains are no longer closely packed in the mixed monolayer and stay phase separated with Alamethicin as evident from the system's high shear rates. Unlike pure DPPC monolayer, the peak positions of q$_{02}$ and q$_{11}$ do not change during flow (Table \ref{T2}). Also, the tilt angle remains fixed with increasing $\sigma$. Strikingly Alamethicin helix coherence length increases with $\sigma$ suggesting that the Alamethicin domains are merging to bigger size promoting more separation of phases in the system.

As noted in Ref. [\cite{kjaer1994some}], the Langmuir films are 2D powders of randomly oriented 2D crystallites in the plane. Bragg reflections do not capture the motion of the crystallites (whenever the reflecting plane satisfies the Bragg condition, it contributes to the Bragg peak). In a way, the motion of the crystallites in a circular streamline path rather helps us to get the powder diffraction pattern. Effectively, the scan is not at a fixed position on the sample but rather the pattern is averaged over large number of crystallites passing through the x-ray footprint.

\begin{figure}
\centering
\includegraphics[width=1.0\textwidth]{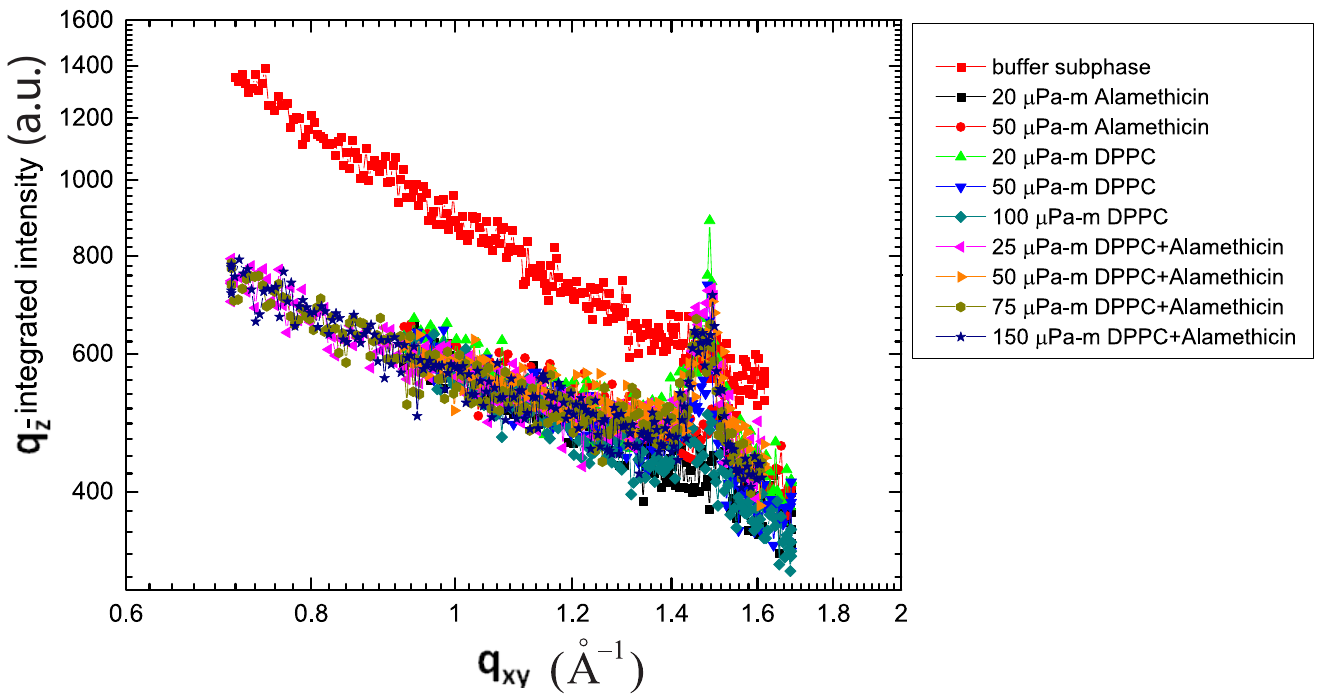}
\caption{$q_z$-integrated intensity vs $q_{xy}$ plot for the monolayers during creep flow. The diffraction data from the clean buffer-subphase surface is also shown.}
\label{F7}
\end{figure}

\begin{figure}
\centering
\includegraphics[width=0.5\textwidth]{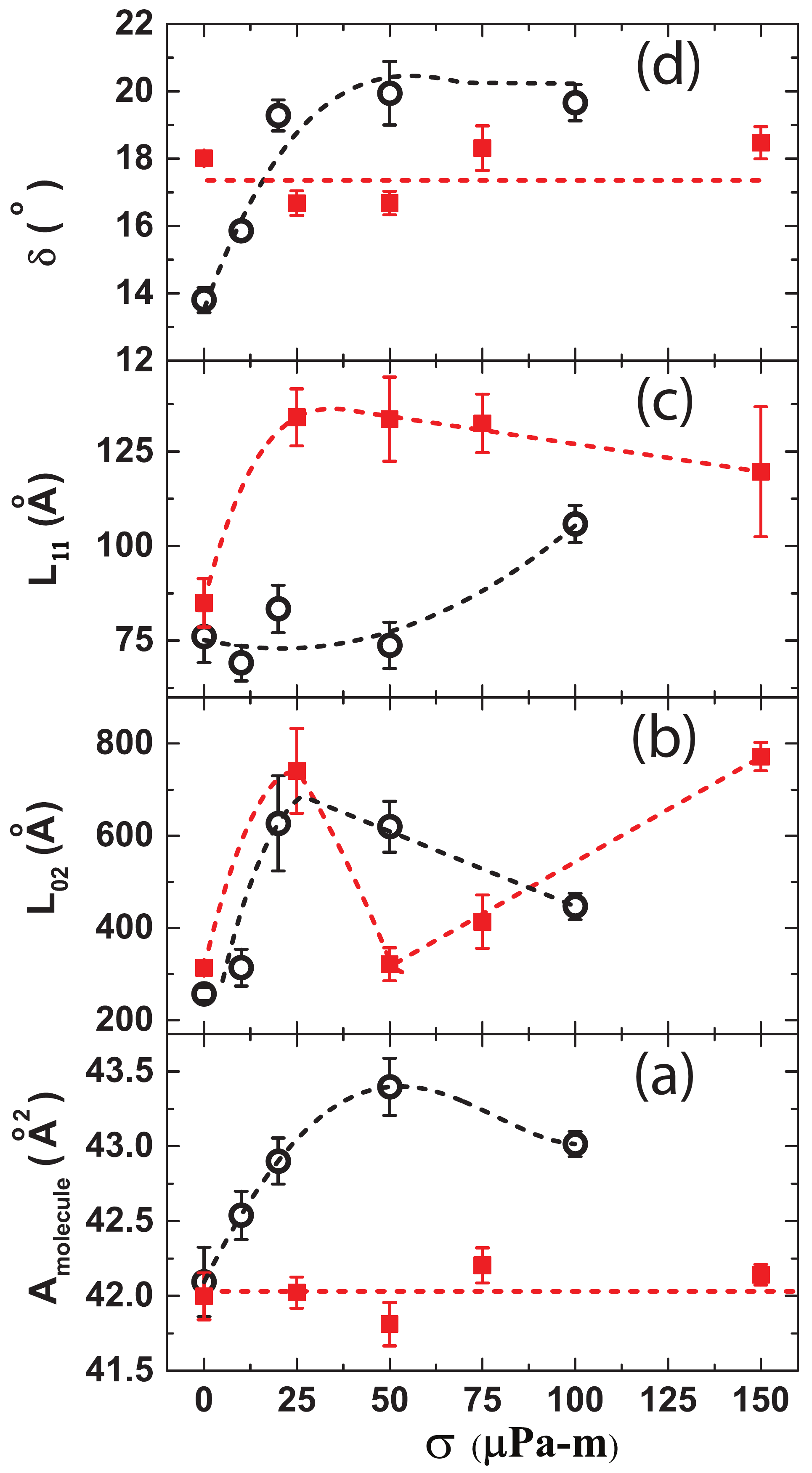}
\caption{(Color online) Area/molecule of DPPC (A$_\mathrm{molecule}$), coherence length corresponding to two DPPC Bragg peaks L$_{02}$; L$_{11}$, tilt angle of DPPC chains ($\delta$) for pure DPPC (open circles) and DPPC-Alamethicin mixed (red squares) monolayers are plotted against $\sigma$. Dotted curves are guides to the eyes.}
\label{F8}
\end{figure}

Figure \ref{F7} shows the log-log plot of $q_z$-integrated intensity vs $q_{xy}$ which decays linearly and confirms the flatness of the interface \cite{sanyal1991x} during the GIXD measurements. For comparison, we have plotted area/molecule (A$_\mathrm{molecule}$), coherence lengths (L$_{hk}$) and the tilt angle ($\delta$) of DPPC for pure and mixed systems (Figure \ref{F8}). For pure DPPC monolayer, the area/molecule (Figure \ref{F8}a) increases rapidly with $\sigma$ and saturates at high values, whereas for mixed monolayer it does not change with $\sigma$. For both the systems, the coherence lengths in [02] direction (L$_{02}$) increases with the increasing shear rate. The data for 25 $\mu$Pa-m of the mixed monolayer does not follow the trend. We propose that the high value of L$_{02}$ corresponding to 25 $\mu$Pa-m is due to the flow merging of crystalline domains during flow jamming transition. On the other hand, the coherence length in [11] direction (L$_{11}$) has a slow increment for the pure DPPC monolayer but shows high value for the mixed monolayer with increasing shear rate. For the pure DPPC monolayer, the tilt angle increases and then saturates at higher $\sigma$, but for the mixed monolayer, it is $\sim$ 17.5 $^{\circ}$ for all $\sigma$.

\section*{Conclusions}
We have described the methodology of Rheo-GIXD, an extension of the well established GIXD technique to study molecular structure under steady shear on the interface by combining interfacial rheology and GIXD. We have demonstrated that the GIXD signal can be captured even when interfacial molecular crystallites move under shear. At low $\sigma$, pure Alamethicin as well as mixed monolayer show jamming behavior after about $\sim$ 100 sec. For a given $\sigma$, the observed steady-state shear rate for the Alamethicin free system is very high confirming the finite flow of the system, but contrary happens for Alamethicin monolayer and the mixed monolayer. Before entering to jammed state, the system flows with a finite shear rate (compare 20 $\mu$ Pa-m data in Figure \ref{F4} and Figure \ref{F5}) which is sufficiently high to be detected by a commercial rheometer. Thus, we can safely conclude that we have a high signal to noise ratio, and this is a genuine flow jamming behavior.

The pure DPPC monolayer shows measurable changes in the lattice parameters and in the tilt angle of the hydrophobic chains. The presence of buffer sub-phase stabilizes the peptide at the air-water interface but does not lead to the binding of the peptide with DPPC head group, as inferred from the observation that, the scattering signal is almost similar in both cases (pure DPPC and DPPC-Alamethicin mixed). The phase separation and the barrel-stave aggregation of an amphipathic peptide in a peptide-lipid matrix in equilibrium \cite{pieta2012direct} are also consistent with our Rheo-GIXD observations under shear. We have shown that the 2D crystallites grow bigger by merging crystalline domains under shear. The structural properties of hexameric pores could not be probed here due to high direct-beam leakage in low q$_{xy}$ region.

Further work along with x-ray reflectivity study on this system will allow us to study the dependence of structural parameters on the velocity gradient. In the future, the underlying transient dynamics will be probed along with the 1D pinhole detector or with the 2D detector. Also, this technique can be used to probe the molecular dynamics near the nonequilibrium phase transition of monolayers under oscillatory shear deformation \cite{bera2019experimental}. We believe that our results will provide motivation for studying the molecular-level structure of many other membranes in non-equilibrium conditions.

\textbf{acknowledgement}

A.K.S. thanks Department of Science and Technology (DST), India for the support through Year of Science Professorship. M.K.S. acknowledges the support Raja Ramanna Fellowship of Department of Atomic Energy (DAE). R. K. thanks DST for the Ramanujan Fellowship. A.K.K. and P.K.B. thank University Grants Commission (UGC) for the D.S.Kothari fellowship and Senior Research Fellowship, respectively. We thank DST for financial assistance through CEFIPRA-SOLEIL-Synchrotron Programme (20140232, AP14/15) to use the Synchrotron Beamtime. We acknowledge SOLEIL for provision of synchrotron radiation facilities and we thank N. Aubert for assistance in using the beamline SIRIUS. We thank Prof. Jean Daillant for fruitful discussions.


\begin{table}
\center\caption{Structural packing parameters of DPPC monolayer for different $\sigma$.}
\begin{tabular}{| c | c | c | c | c | c |}
\hline 
\rule{0pt}{3ex}$\sigma$ & d-spacings & unit cell & A$_\mathrm{molecule}$ & Coherence & Tilt angle\\
$[\mu$Pa-m$]$	& $[$\AA$]$ & dimensions $[$\AA$]$ & $[$\AA$^2]$ & length $[$\AA$]$ & $\delta$ $[^{\circ}]$ \\
\hline
0 & d$_{11}=4.293\pm0.015$ & a$=4.986\pm0.024$ & $42.09\pm0.23$ & L$_{11}=76\pm7$ & $13.8\pm0.4$ \\
	& d$_{02}=4.221\pm0.003$ & b$=8.443\pm0.006$ &  & L$_{02}=256\pm15$ &  \\
\hline
10 & d$_{11}=4.328\pm0.010$ & a$=5.042\pm0.017$ & $42.54\pm0.16$ & L$_{11}=69\pm5$ & $15.9\pm0.3$ \\
	& d$_{02}=4.218\pm0.002$ & b$=8.437\pm0.004$ &  & L$_{02}=314\pm40$ &  \\
\hline
20 & d$_{11}=4.350\pm0.009$ & a$=5.072\pm0.015$ & $42.90\pm0.15$ & L$_{11}=83\pm6$ & $19.3\pm0.5$ \\
	& d$_{02}=4.229\pm0.002$ & b$=8.459\pm0.005$ &  & L$_{02}=627\pm103$ &  \\
\hline
50 & d$_{11}=4.384\pm0.012$ & a$=5.123\pm0.020$ & $43.40\pm0.19$ & L$_{11}=74\pm6$ & $19.9\pm0.9$ \\
	& d$_{02}=4.236\pm0.002$ & b$=8.472\pm0.004$ &  & L$_{02}=620\pm55$ &  \\
\hline
100 & d$_{11}=4.359\pm0.005$ & a$=5.088\pm0.008$ & $43.02\pm0.08$ & L$_{11}=106\pm5$ & $19.7\pm0.5$ \\
	& d$_{02}=4.227\pm0.002$ & b$=8.455\pm0.003$ &  & L$_{02}=447\pm29$ &  \\
\hline
\end{tabular}
\label{T1}
\end{table}

\clearpage

\begin{table}
\center\caption{Structural packing parameters of DPPC-Alamethicin mixed monolayer for different $\sigma$.}
\begin{tabular}{| c | c | c | c | c | c |}
\hline 
\rule{0pt}{3ex}$\sigma$ & DPPC d-spacings, & DPPC unit & DPPC & Coherence length & Tilt angle\\
$[\mu$Pa-m$]$ & Alamethicin pitch & cell dimensions & A$_\mathrm{molecule}$ & $[$\AA$]$; DPPC L$_{hk}$, & $\delta$ $[^{\circ}]$ \\
	& $[$\AA$]$ & $[$\AA$]$ & $[$\AA$^2]$ & Alamethicin L$_p$ &  \\
\hline
0 & d$_{11}=4.296\pm0.009$ & a$=4.999\pm0.016$ & $42.00\pm0.16$ & L$_{11}=85\pm6$ & $18.0\pm0.2$ \\
	& d$_{02}=4.201\pm0.003$ & b$=8.402\pm0.005$ &  & L$_{02}=314\pm17$ &  \\
	& p$=4.160\pm0.005$ &  &  & L$_p=396\pm96$ &  \\
\hline
25 & d$_{11}=4.294\pm0.005$ & a$=4.993\pm0.008$ & $42.02\pm0.10$ & L$_{11}=134\pm8$ & $16.7\pm0.4$ \\
	 & d$_{02}=4.208\pm0.003$ & b$=8.416\pm0.007$ &  & L$_{02}=741\pm92$ &  \\
 	 & p$=4.150\pm0.005$ &  &  & L$_p=1510\pm459$ &  \\
\hline
50 & d$_{11}=4.284\pm0.007$ & a$=4.981\pm0.012$ & $41.81\pm0.15$ & L$_{11}=134\pm11$ & $16.7\pm0.4$ \\
	 & d$_{02}=4.197\pm0.004$ & b$=8.395\pm0.009$ &  & L$_{02}=321\pm36$ &  \\
   & p$=4.140\pm0.003$ &  &  & L$_p=1611\pm305$ &  \\
\hline
75 & d$_{11}=4.303\pm0.005$ & a$=5.003\pm0.009$ & $42.20\pm0.12$ & L$_{11}=132\pm8$ & $18.3\pm0.7$ \\
	 & d$_{02}=4.218\pm0.004$ & b$=8.435\pm0.008$ &  & L$_{02}=413\pm58$ &  \\
	 & p$=4.153\pm0.006$ &  &  & L$_p=1250\pm355$ &  \\
\hline
150 & d$_{11}=4.300\pm0.005$ & a$=4.999\pm0.008$ & $42.14\pm0.07$ & L$_{11}=120\pm17$ & $18.5\pm0.5$ \\
	 & d$_{02}=4.215\pm0.001$ & b$=8.429\pm0.001$ &  & L$_{02}=772\pm31$ &  \\
	 & p$=4.151\pm0.002$ &  &  & L$_p=1050\pm91$ &  \\
\hline
\end{tabular}
\label{T2}
\end{table}

\clearpage

\textbf{SUPPORTING INFORMATION}

\section*{Section A}
\begin{figure}[hb]
\centering
\includegraphics[width=0.5\textwidth]{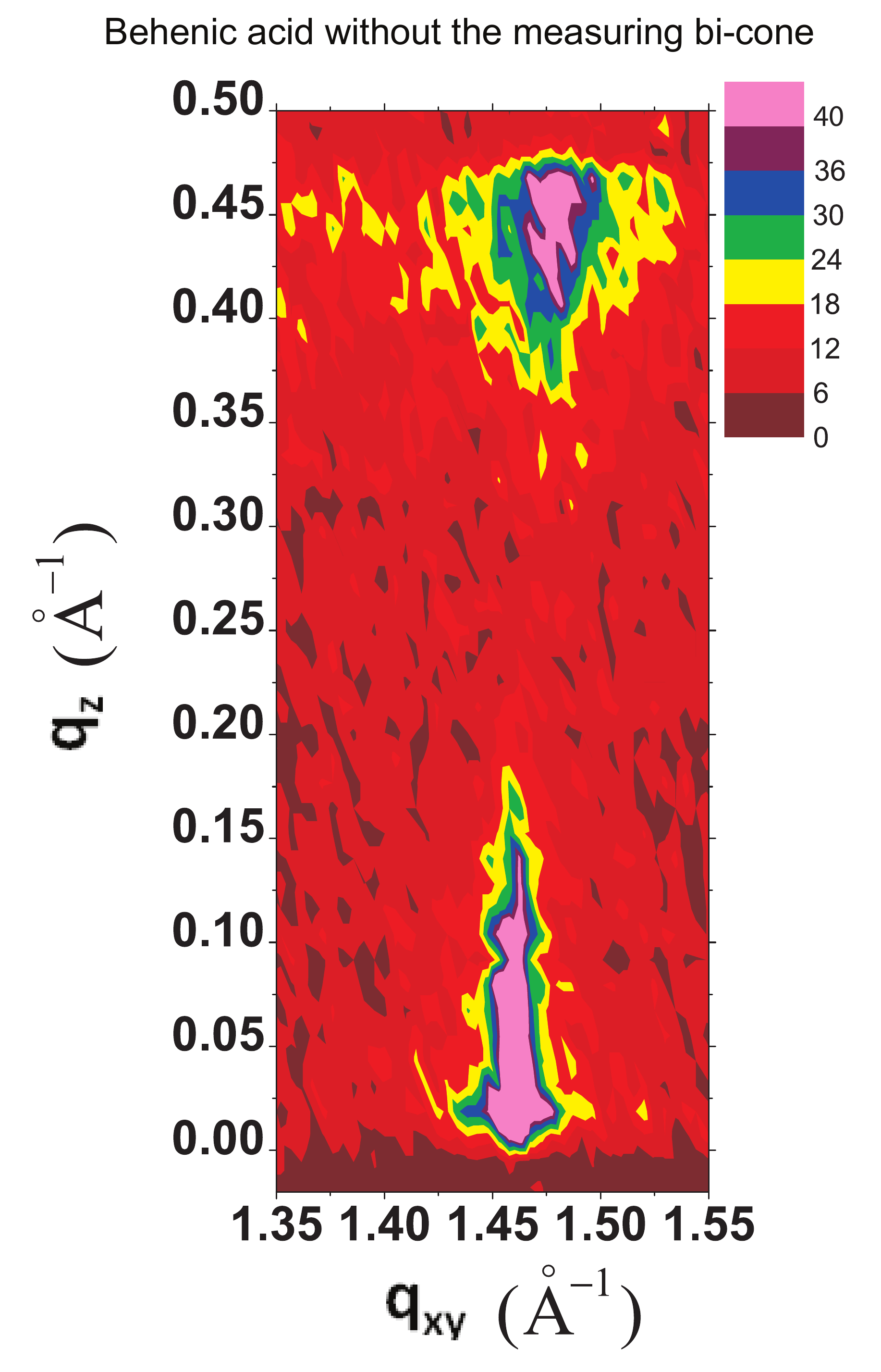}
\caption{(Color online) GIXD pattern from the Behenic acid monolayer resting on the water surface in the cell.}
\label{Behenic2D}
\end{figure}

We have used the Behenic acid as a test sample to calibrate our GIXD setup. We have water filled the cell without the bicone at its measuring position and spread the Behenic acid monolayer to do the GIXD scan. The 2D diffraction plot (Figure \ref{Behenic2D}) is matching with literature (article ref. 24).

\clearpage

\section*{Section B}
\begin{figure}
\centering
\includegraphics[width=\textwidth]{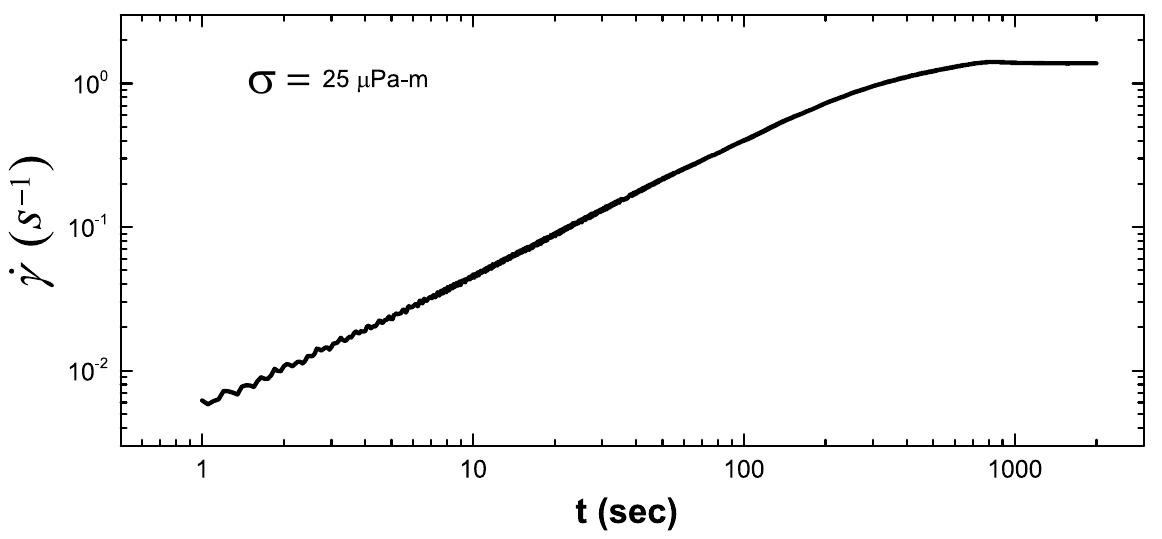}
\caption{(Color online) Creep flow of cholesterol monolayer with surface concentration 20 \AA$^2$/molecule.}
\label{cholesterol}
\end{figure}

\begin{figure}
\centering
\includegraphics[width=\textwidth]{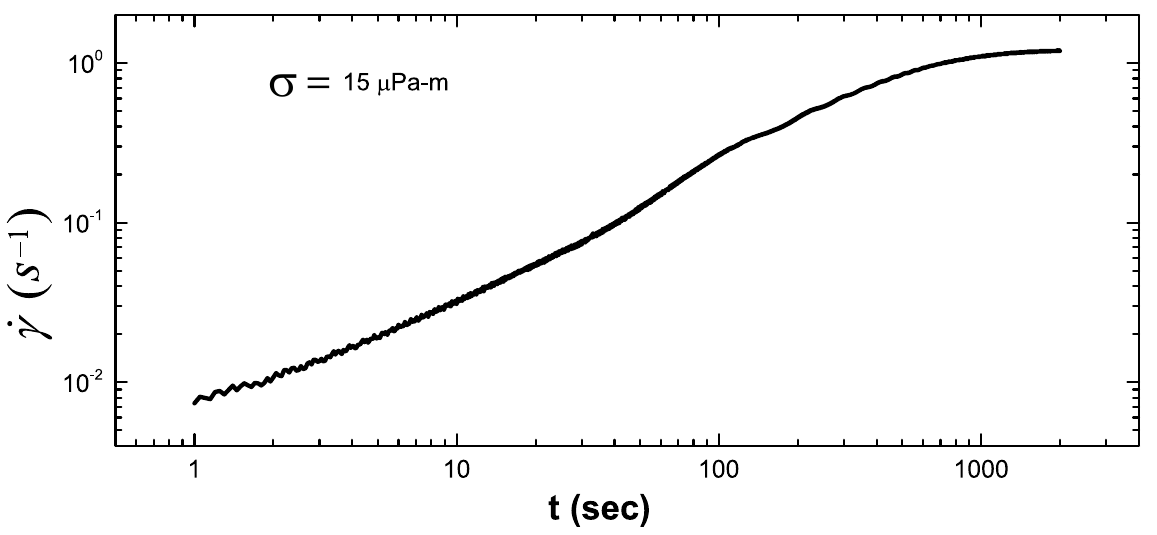}
\caption{(Color online) Creep flow of cholesterol + DPPC monolayer (mole ratio 4:1; surface concentration 20 \AA$^2$/molecule.}
\label{cholesterolDPPC}
\end{figure}

The sensitivity of the monolayers to the very small stress (corresponding to the small torque imparted by the rheometer) was calibrated using two highly concentrated test samples (i) cholesterol monolayer and (ii) cholesterol+DPPC mixed monolayer. Cholesterol is known for having very low surface viscosity in monolayer from and also it reduces surface viscosity of monolayers of the other dipalmitoyl phospholipids (article ref. 25). The expected shear rate should be very high even for a very small applied stress. Figure \ref{cholesterol} and Figure \ref{cholesterolDPPC} show the measured shear rate ($\dot{\gamma}$ of the two monolayers during creep flow. The very high values of the shear rate confirm that the torque applied by the rheometer is sufficient enough for the molecularly thin Langmuir films to flow at a finite shear rate.)

\end{document}